\documentclass[10pt]{article}

\usepackage{amsmath}
\usepackage{amssymb}

\usepackage{cite}

\usepackage{hyperref}

\usepackage{lineno}

\usepackage{microtype}
\DisableLigatures[f]{encoding = *, family = * }

\usepackage[labelfont=bf,labelsep=period,justification=raggedright]{caption}

\makeatletter
\renewcommand{\@biblabel}[1]{\quad#1.}
\makeatother

\usepackage{graphicx}
\usepackage{float}

\date{}

\begin{document}

% Title must be 150 characters or less
\begin{flushleft}
{\Large
\textbf{Estimating Food Consumption and Poverty indices with Mobile Phone Data}
}
% Insert Author names, affiliations and corresponding author email.
\\
Adeline Decuyper$^{1}$, 
Alex Rutherford$^{2}$, 
Amit Wadhwa$^{3}$
Jean Martin Bauer$^{3}$
Gautier Krings$^{1, 4}$
Thoralf Gutierrez$^{4}$
Vincent D. Blondel$^{1}$
Miguel A. Luengo-Oroz$^{2, \ast}$
\\
\bf{1} ICTEAM institute, Universit\'e catholique de Louvain, Louvain-La-Neuve, Belgium
\\
\bf{2} United Nations Global Pulse, New York, United States
\\
\bf{3} Vulnerability Analysis and Mapping, World Food Programme, Rome, Italy
\\
\bf{4} Real Impact Analytics, Brussels, Belgium
\\
$\ast$ E-mail: miguel@unglobalpulse.org (MLO)
\end{flushleft}

% Please keep the abstract between 250 and 300 words
\section*{Abstract}
Recent studies have shown the value of mobile phone data to tackle problems related to economic development and humanitarian action. In this research, we assess the suitability of indicators derived from mobile phone data as a proxy for food security indicators. We compare the measures extracted from call detail records and airtime credit purchases to the results of a nationwide household survey conducted at the same time. Results show high correlations ($>.8$) between mobile phone data derived indicators and several relevant food security variables such as expenditure on food or vegetable consumption. This correspondence suggests that, in the future, proxies derived from mobile phone data could be used to provide valuable up-to-date operational information on food security throughout low and middle income countries.

\section*{Introduction}
In recent years the explosion in the use of digital services has led to what has come to be called the ``Big Data Revolution". This revolution has fundamentally changed how companies are able to understand their users through analysis of data produced passively. However, more recently, as low-cost mobile handsets and internet usage have proliferated in developing countries, reaching $90\%$ coverage in the developing world in 2014 \cite{ITU}, the potential of these signals to transform development and humanitarian action has emerged.

Big Data has the potential to guide policy makers by providing an alternative to traditional data sources such as costly and time-consuming manual surveys \cite{deville2014dynamic}. For planning purposes, access to reliable and up-to-date statistical data is essential to humanitarian organizations when deciding where and when help is most needed. Many rich data sources exist with the promise of providing early warning and real-time monitoring of vulnerable populations including remote sensing, social media, remittances and anonymized mobile phone records.

In particular, many recent studies have shown how valuable mobile phone data in the form of Call Detail Records (CDRs) can be used to guide agile development policy and humanitarian action \cite{D4Db,BCD13,LDP13}. The Orange D4D challenge \cite{D4Db}, using C\^{o}te d'Ivoire as a case study, led to a large number of innovative uses of CDR data ranging from mobility modeling for transport optimization \cite{BCD13} to epidemic modeling \cite{LDP13} and network analysis of social communities \cite{AKK14}. The challenge was such a success that other challenges of big data exploration have followed \cite{tel2014,demontjoye2014d4d}, and the results of a second development challenge, this time on data from Senegal, will be presented in the course of 2015 \cite{demontjoye2014d4d}. 

Given the importance of mobile devices to people in developing economies for accessing information and economic opportunities, phone usage data represents a clear barometer of a user's socio-economic conditions in the absence or difficulty of collecting official statistics \cite{deville2014dynamic, FSVF13, GKB13}. 

Eagle \textit{et. al.} have addressed this question by analyzing the mobile and landline network of a large proportion of the population in the UK. They introduced the \textit{Social Diversity} measure, that quantifies how equally communication time is shared amongst one's contacts \cite{EMC10}. It was shown that the social diversity is a good proxy for the variation of the poverty levels within the population of the UK. 

In a related study, Frias-Martinez \textit{et. al.} have analyzed the relationship between mobility observed from cell phone traces and the socio-economic levels of the different areas of a city in Latin-America \cite{FV12}. They showed which mobility indicators correlated best with socio-economic levels, and introduced in a further paper \cite{FSVF13} a method to predict the socio-economic level of an area based on mobile phone traces of users. 

As part of the D4D challenge, Smith-Clarke \textit{et. al.} have compared several features extracted from cell-tower-aggregated volumes of calls to poverty levels in C\^{o}te d'Ivoire \cite{SMC14}. They showed a good correspondence between several features of users' mobile phone usage and available data on poverty levels. However, the poverty data was only available at a very coarse spatial resolution and was several years older than the mobile phone dataset. Ironically, the absence of appropriate `ground-truth' data from surveys and census on which to validate novel data sources can often limit the applicability of studies using CDRs.

In a recent paper, Gutierrez \textit{et. al.} also presented a few hypotheses addressing the question of mapping poverty with mobile phone data, adding an analysis of airtime credit purchases in C\^{o}te d'Ivoire \cite{GKB13}. 
In a different approach, rather than studying whole regions at a time, Blumenstock recently studied in more details the relationship between \textit{individual} indicators of wealth and the history of mobile phone transactions \cite{blumenstock2014calling}. The study indicates preliminary evidence of a relationship between socioeconomic status and mobile phone use, at the individual level. These different approaches are different, yet strongly related and complementary.
To our knowledge, however, no study has yet had the opportunity to analyze airtime credit purchases and their relationship with ground truth data on food security gathered from wide scale household surveys covering a whole country.  

The objective of this research is to assess the suitability of metrics derived from mobile phone data, specifically CDRs and airtime credit purchases, as proxies for food security and poverty indicators in a low-income country context. We use a country of central Africa as a case study, and compare features measured from mobile phone data collected in 2012 with the results of a detailed food security survey conducted at the same time. The World Food Programme regularly assesses the situation of nutrition, their reports with updated information are available online \cite{WFP12}. While food access has greatly improved between 2009 and 2012, there is still approximately one household in five that is food insecure in these regions. One of the recommended steps for the way forward to tackle malnutrition in Africa is to obtain better and more frequent information on food security. Therefore, a real-time proxy of relevant trends on food security indicators would be of great interest for regions where household surveys can be time and resource consuming if they are possible at all. In order to complement the information obtained from  food security indicators, we also assess the possibility of using the same mobile phone data to map poverty levels, using the results of a survey on non-monetary poverty as a ground truth comparison.

% You may title this section "Methods" or "Models". 
% "Models" is not a valid title for PLoS ONE authors. However, PLoS ONE
% authors may use "Analysis" 
\section*{Materials and Methods}
\subsection*{Data Description}
\subsubsection*{CDR data and Airtime credit purchases}
The Call Detail Records we used for this study include seven months of mobile phone activity of a significant portion of the population in the country of interest, from one large mobile phone carrier. Each CDR contains the following information: caller ID, callee ID, cell tower initiating the call, date and time. In addition, we also analyze the history of six months of airtime credit purchases, recorded at the same time as the CDRs, and representing the same population of users. Each top-up in the data contains the user ID, top-up amount, date and time of top-up.   

\subsubsection*{Food consumption survey}
To assess the suitability of mobile phone data as a proxy for food security indicators, we use the results of a survey conducted in 2012 reaching 7500 households spread geographically across the country so as to be representative. The results include all answers to the survey questions at the household level, and other computed measures of food consumption indicators.  

The questionnaires contained 486 questions including demographics of the household members, characteristics of the home shelter and household income. There were many specific questions related to food access and consumption, including the number of times each type of food was consumed in the last 7 days, how long did the stocks from each season last, difficulties in accessing food throughout the year among others.

The data also includes some composite measures from several questions. All in all there were 1018 variables in the data for each of the 7500 interviewed households.
Relevant composite measures are the \textit{Food Consumption Score} (FCS) \cite{FCS08} and the \textit{Coping Strategy Index} (CSI) \cite{CSI08}. The FCS is computed weighting the consumption per household of several food items over the last week. The FCS value is typically thresholded to classify the households into poor, borderline and acceptably food secure. The CSI takes into account how, and how often the household will react when it does not have enough available food or money to buy food. The CSI is thus a measure of the severity and frequency of the coping strategies that a household uses when its access to food is restricted.

\subsubsection*{Poverty indices}
In this analysis, we also use poverty indices available from the results of a national survey conducted in 2012 to assess the levels of non-monetary poverty. 
Non-monetary poverty indices measure the deprivation or the difficulty of access to certain goods and services that are necessary for everyone.  
We compare the CDR metrics with three indicators of poverty: the percentage of people that are poor in each area, the intensity of deprivation among the poor and the Multidimensional Poverty Index (MPI) which takes into account three dimensions of equal weight: education, health and living standards. The MPI is computed as a product of the two previous variables. 

\subsection*{Methodology}
In our analyses, we compare measures extracted from mobile phone data, and from the food consumption survey aggregated by geographical areas (hereafter referred to as \textit{``sectors"}), which cover a population around 10 000 - 50 000 habitants depending on the rural or urban nature. 
  
\subsubsection*{Mobile phone data processing}
First, a home location was assigned to each user. To do this, we selected as home location the cell-tower from which the user has made the most calls between 6pm and 8am \cite{TBD14}. For comparison, we repeated the same analyses when assigning people to a home location where they had made the most calls at any time of day or night. As expected, results showed a better correspondence with survey variables when using the night-time calls only.\\ 

For each user, a series of measures characterizing user top-up behavior and network structure were computed. For top-up behavior, the average top-up, the sum of top-ups over the six month period, and the minimum and maximum values of top-ups were all computed. For the network structure, the social diversity as introduced in \cite{EMC10} was measured. For each sector, aggregated measures of the users' characteristics: mean, median, standard deviation, coefficient of variation were derived. 

\subsubsection*{Food consumption survey data processing}
A set of numerical metrics related to food security were considered, against which the relation with CDR variables could be investigated. These variables are divided into three categories: household characteristics (V1), food consumption variables (V2), and wealth variables (V3).
The household characteristics include variables such as the size of the household, the crowding index or the age of the different members of the household.  
The food consumption variables represent the quantity or frequency of household food access, as well as aggregated measures of food consumption such as the FCS. 
The wealth characteristics include variables such as the monthly expenses on a range of items including food and non-food categories, the proportion of expenses spent on different items, and different measures of the income of the household and total expenses. 
Each of those variables is aggregated by sector by taking the mean over all households that belong to that sector. 

\subsubsection*{Comparing food consumption and mobile phone data}
For both the food consumption survey and the mobile phone data, there is a set of indicators aggregated by sector. 
A large correlation matrix was then computed comparing 13 mobile phone variables to 232 food consumption and poverty indicators, to investigate which of these indicators could be estimated with mobile phone data, and which could not. Each pair of one mobile phone variable and one food consumption variable was tested for linear correlation as determined by the Pearson correlation coefficient. Associated p-values and confidence intervals were also computed.  

In order to bootstrap the recorded correlations, the same correlations were calculated using sectors in a shuffled order as a null hypothesis, and it was observed that the correlations dropped dramatically. The highest correlations in that case were only 0.3 and the variables that correlated highly were not the same as when comparing the sectors' measures in the right order, and also changed between different random trials. 

Additionally, we used linear and symbolic regression to fit a function of the mobile phone variables to the levels of several food indicators that were of particular interest (such as, for example, the Food Consumption Scores, or total expenditure on food).

% Results and Discussion can be combined.
\section*{Results}
\begin{figure}[H]
\includegraphics[scale=0.43]{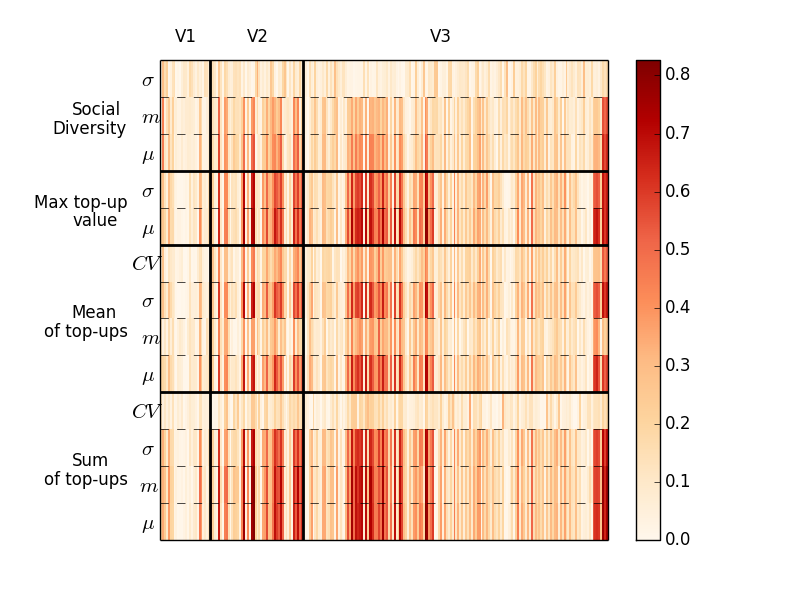}
\caption{{\bf Absolute value of pearson correlation coefficients between mobile phone features and food security measures. } On the x-axis: V1: household characteristics (26 variables); V2: food consumption metrics (48 variables); V3: Expenses metrics (158 variables). On the y-axis: sector aggregated measures (mean $\mu$, median $m$, standard deviation $\sigma$ and coefficient of variation $CV$) of each customer's sum, mean, max value of top ups and social diversity. Absolute values are shown for easy visual comparison of classes of features which correlate strongly, the sign of all significant correlations are intuitive i.e. CDR features indicating greater expenditure are matched by survey responses indicating greater food security.}
\label{correl}
\end{figure}

Correlations between mobile phone metrics and food consumption variables are depicted in figure \ref{correl}, and some detailed values are available in tables \ref{table_food} and \ref{table_cor}.
Very high correlations ($>0.7$) were found between several food security indicators and measures computed from airtime purchases. Generally, the sum of expenses and the maximum top-up value are the mobile variables that correlate best with several of the food indicators. On the other hand, the coefficient of variation of the sum and average of top-ups did not result in any high correlation.\\

The highest correlation value of all the models tested in this study corresponds to a quadratic model with the sum of expenses and the average amount of top-up as mobile phone derived indicators that shadows the survey variable measuring the amount of food expenses (0.89). A linear model of mobile variables correlated at 0.69 with the food consumption score (FCS). 
The Coping Strategy Index (CSI), however, only showed a correlation of 0.25 with the mobile phone variables, not significant enough to draw conclusions of a clear relationship between mobile phone consumption and the CSI. 
As the survey variables are hardly independent from one another \cite{WFP12}, when a mobile phone metric correlates highly with one of the survey variable, it also presents a strong correlation with several others. 
The same reasoning is also valid for mobile phone variables. The average of the sum of top-ups is hardly independent from the median of the sum of top-ups. Observations also confirmed that when one of the top-up variables correlates well with survey variables, then so do also many other top-up metrics.\\

Looking in detail at the relations between the consumption of each food item (survey question ``how many times have you eaten [item] in the last 7 days?") and mobile phone expenditure (measured by the average of the sum of airtime expenses per user over 6 months), we found different correlation values ranging from no relation to very high correlation between certain food items and airtime expenses. See results in table \ref{table_food}. In particular the consumption of vitamin rich vegetables, rice, bread, sugar and fresh meat have a very high correlation ($>0.7$) with the airtime purchases. On the other side of the spectrum, broadly cultivated items like cassava and beans have no relation with the expenditure on mobile phones. These results are particularly interesting, as the products from the first group are mainly bought in the market while the products with consumption uncorrelated to top up values are cultivated for personal consumption. Therefore this study is compatible with a new hypothesis: expenditure in mobile phone top up is proportional to the expenditure in food in the markets.\\

An additional interesting observation is that the correlation between airtime expenses and white sweet potato consumption is the only one with a significant negative correlation. The more money spent on mobile phone expenses, the less days per week people eat sweet potato, a very cheap and broadly cultivated item. This suggests that when people can afford to, they will reduce their consumption of sweet potato. This observation is also coherent with a negative correlation with the share of food consumed in the households obtained from own production ($-0.64$).\\

As could be expected, other high correlations were observed between mobile phone expenses, and expenses of households on other non-food related items. In particular, expenses in education, school fees and uniforms had a 0.72 correlation. Reported expenses in communication and top up expenses had a 0.69 correlation (note that this question in the household survey refers broadly to communication and not just to mobile expenses). \\

When analyzing the relation between mobile phone variables and poverty indices, we have been able to create a proxy indicator for multidimensional poverty index at the sector level ($>.8$ correlation). Interestingly, we have found that the most important variable to shadow poverty levels is the top up information. Regarding the distribution of the \textit{denomination} of credit added by users in different sectors; this was fairly stable, invariably dominated by identical amounts which corresponds with the most common fixed denominations available for physical purchase. Therefore, 10 small credits of size X would be used in place of one larger denomination of size 10X. \\

% We only support three levels of headings, please do not create a heading level below \subsubsection.
\section*{Discussion}
This study is a first step towards demonstrating the utility of mobile phone data, as a novel new data source for food security monitoring. In this research, mobile phone data including Call Detail Records and airtime credit purchases were used to investigate the relationship between mobile phone use and food security indicators. Mobile phone derived indicators were compared to the variables of an in depth household level survey conducted at the same time as the mobile phone data was collected.\\

Results showed that top up derived metrics are significantly correlated with composed food security indicators and with the consumption of selected food items such as vitamin rich vegetables, rice, bread, sugar and fresh meat. Therefore this study suggests and supports the new hypothesis that the mobile phone expenditure could be used as a proxy for food expenditures in market dependent households, mostly shadowing the consumption and expenses of items that are usually bought on the market rather than produced directly by the consumer. In this study, households spend about 5-10 \% of the money spent in food on mobile services.\\

This research also shows that poverty levels can be inferred from top up information at quite a granular level in a low income economy. 
In \cite{SMC14}, Smith-Clarke \textit{et. al.} observed good correlations between poverty indices and several indicators derived from mobile phone activity. Our observations confirmed their results that mobile phone activity could be linked to poverty indices. Additionally, our analyses showed that top-up metrics can achieve very good correlations with smaller confidence intervals than activity metrics presented in \cite{SMC14}.
Contrary to the results presented in \cite{EMC10}, results of our analysis showed poor correlations and opposite to those observed by Eagle \textit{et. al.} between social diversity and poverty indices. There are several possible reasons for that discrepancy. First, the socio-economic level indicators used in \cite{EMC10} and the MPI are not exactly the same. Secondly, the interplay of social diversity with socio economic levels could be different in Africa than in the UK. This could be due to very different social norms and structures or the greater degree of economic disparity. And lastly, a possible explanation is that the authors of \cite{EMC10} had access to a database with almost universal coverage of the UK population allowing nearly the entire social graph to be reconstructed, whereas our study included only users of and calls between users of a single carrier. This would suggest that top-up data metrics are a more widely applicable measure of food security and poverty levels, since they do not rely on a full sampling of the call network.\\

\begin{figure}[H]
\includegraphics[scale=0.4]{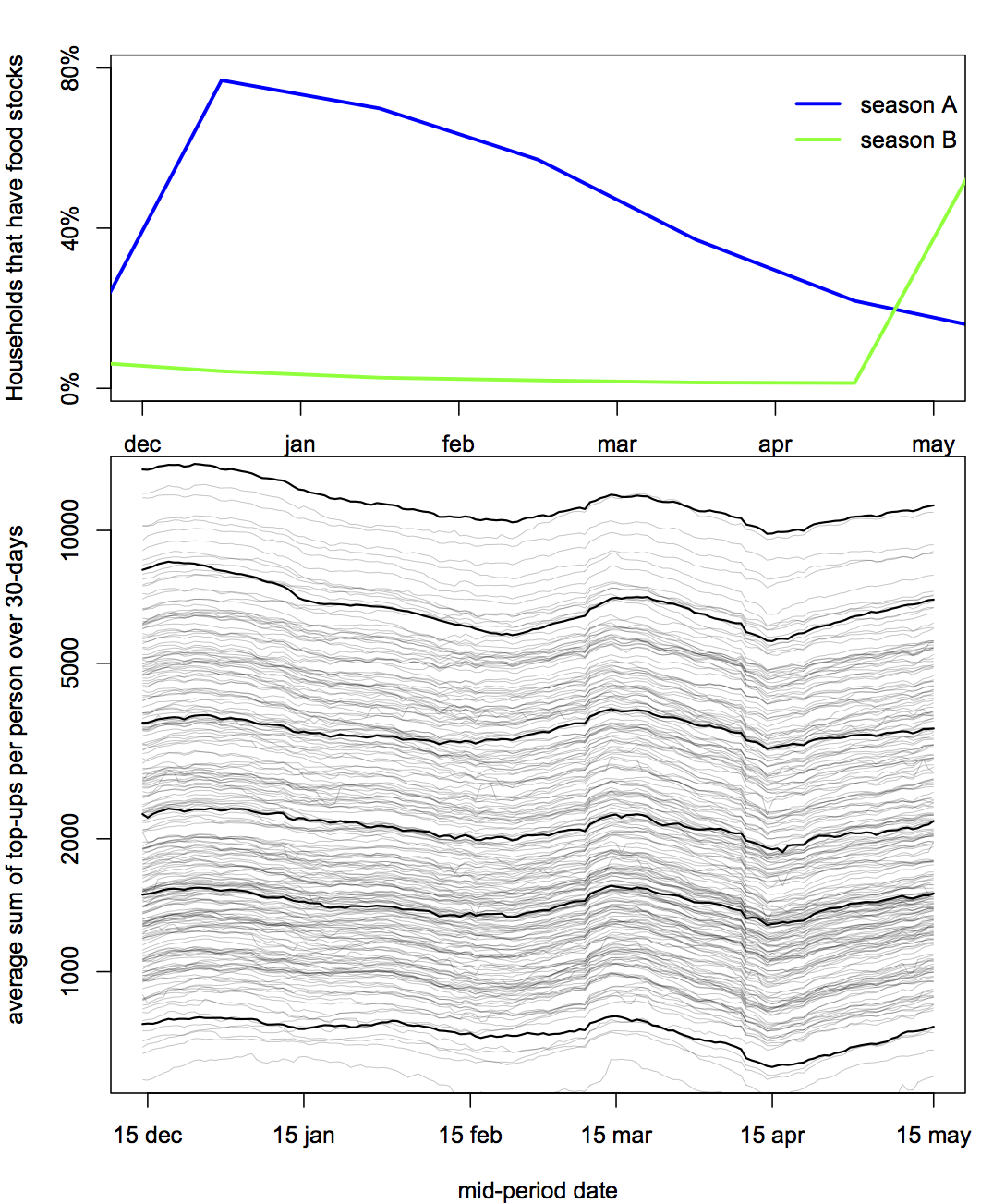}
\caption{{\bf Evolution of food stocks and of mobile phone expenses.} (above) Percentage of households that have food stocks from each season's harvest. (below) One line per sector, evolution of the average sum of expenses over a rolling window of 30 days in logarithmic scale. The date on the x-axis represents the middle of the period, i.e. the label 15 dec. corresponds to the sum of expenses over the period from 1st to 30th of December.}
\label{rolsum30}
\end{figure}

Further methodological developments should be devoted to explore the dynamic evolution of the mobile phone derived indicators in relation to food security assessments and would need several household surveys at different time points as ground truth data. In this research, we did not have access to two different household surveys at two different time points covered by the mobile data. However, as a exploratory experiment, we computed the average sum of expenses for each sector over a rolling time window of 30 days (see figure \ref{rolsum30}), observing that the evolution of the amount spent on mobile phones follows the same general trends across the different geographies.
We then qualitatively compared this evolution with the impact of seasonal changes on food stocks levels in the region of interest (see figure \ref{rolsum30}). Interestingly, both food stock levels and the average of top up per user followed similar dynamics. During periods of high food availability i.e. following a harvest, supply and demand will lead to more economic activity in the markets, which - we hypothesize- is then reflected in the top up derived mobile indicators. This suggests that the changing top-up patterns are responsive to economic conditions on timescales that could potentially provide an early warning for humanitarian action. We might hypothesize that if a specific event leads to stocks in food running out in a specific region, that the shortage in food could trigger a signal in the top-up data.\\

While CDR studies have great potential to provide detailed real-time monitoring of vulnerable populations, the need to consider the risk to individual privacy is extremely pressing. Although CDR data used for research is de-identified, it is still possible that individuals with unique phone activity could be identified from the anonymous dataset \cite{unique}. As a potential solution to such privacy concerns, it has been suggested that mobile users be spatially aggregated so that individuals are not considered. This approach does potentially introduce an associated drop in operations utility of the dataset as detail is lost. However, this study has demonstrated that such an approach can be adopted to provide a real time proxy for traditional information gathering processes at an equivalent spatial resolution. \\

All in all, this analysis shows that simple statistics on top-up data could very accurately provide valuable information about the evolution of food security at a fine grained spatial level. In particular, we have begun to outline which aspects of ground truth food security information are reflected in CDR and mobile phone top-up information. We emphasise that such analyses are not intended to replace traditional in depth surveys, but can provide real-time feedback on programs and interventions between quarterly, biannual or even annual deployments of household or individual level surveys. We also see the potential of using aggregated measures of top-up activity as an early warning system that further information should be gathered or that in-depth surveys should be conducted. This research illustrates a feasible mechanism whereby mobile data could be shared outside on a mobile company at a lower resolution than the individual level, yet still providing information to guide policy.\\

With this in mind, we envision that such a process of real-time monitoring of food security could be operationalized in collaboration with mobile carriers. The information of expenditure is already collected in real-time for billing purposes, and we have demonstrated that potentially sensitive personal data can remain within the mobile company and simply be aggregated to provide an indicator to government and partners to react to sudden changes in food access as measured through the correlation of mobile phone top ups and food expenditure.

% Do NOT remove this, even if you are not including acknowledgments.
\section*{Acknowledgments}
This research was made possible with the support of Real Impact Analytics. We would also like to thank the Global Pulse team in New York and the Vulnerability Analysis and Mapping team from WFP. In particular, we would like to thank Olivia De Backer for her valuable help in the project preparation and overall coordination. \\

%\section*{References}

% Either type in your references using
% \begin{thebibliography}{}
% \bibitem{}
% Text
% \end{thebibliography}
%
% OR
%
% Compile your BiBTeX database using our plos2009.bst
% style file and paste the contents of your .bbl file
% here.
% 

%\bibliography{bib_RIGP}
%\bibliographystyle{plos2009}

\newpage
\section*{Figures}
% This section is for figure legends only, do not include
% graphics in your manuscript file.
%
%\begin{figure}
%\caption{
%{\bf Bold the first sentence.}  Rest of figure caption.  
%}
%\label{Figure_label}
%\end{figure}

\begin{figure}[H]
\includegraphics[scale=0.6]{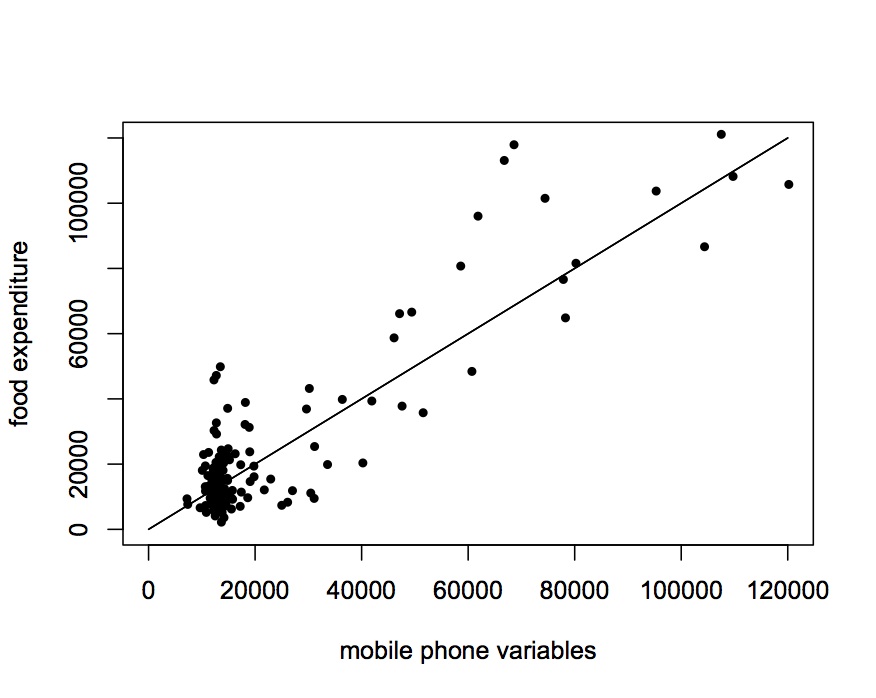}
\caption{{\bf Quadratic combination of CDR variables against expenses on food.} Correlation coefficient: 0.89.}
\label{combili_Foodexp}
\end{figure}

\begin{figure}[H]
\includegraphics[scale=0.6]{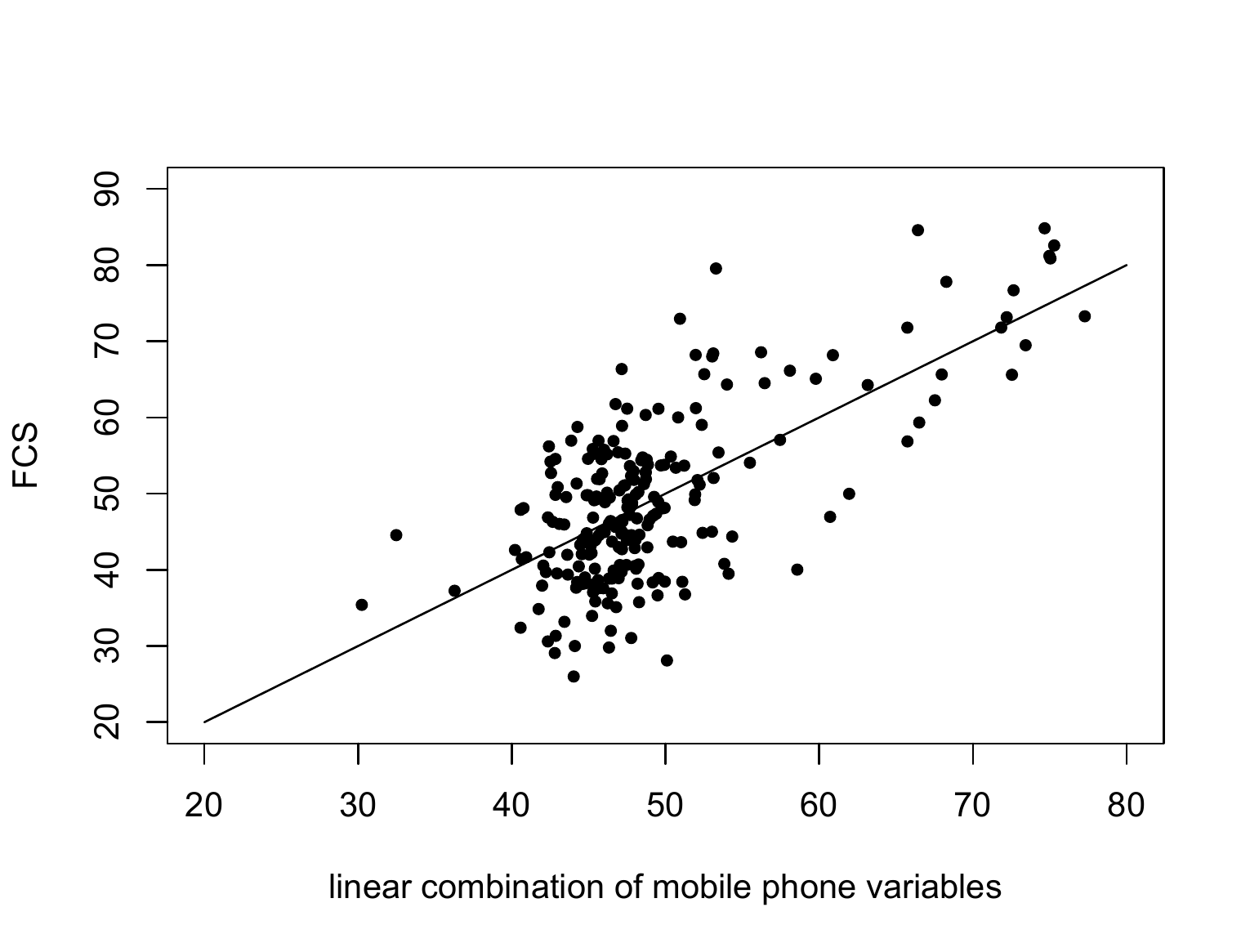}
\caption{{\bf Linear combination of CDR variables against FCS.} Correlation coefficient: 0.69.}
\label{combili_FCS}
\end{figure}

\begin{figure}[H]
\includegraphics[scale=0.6]{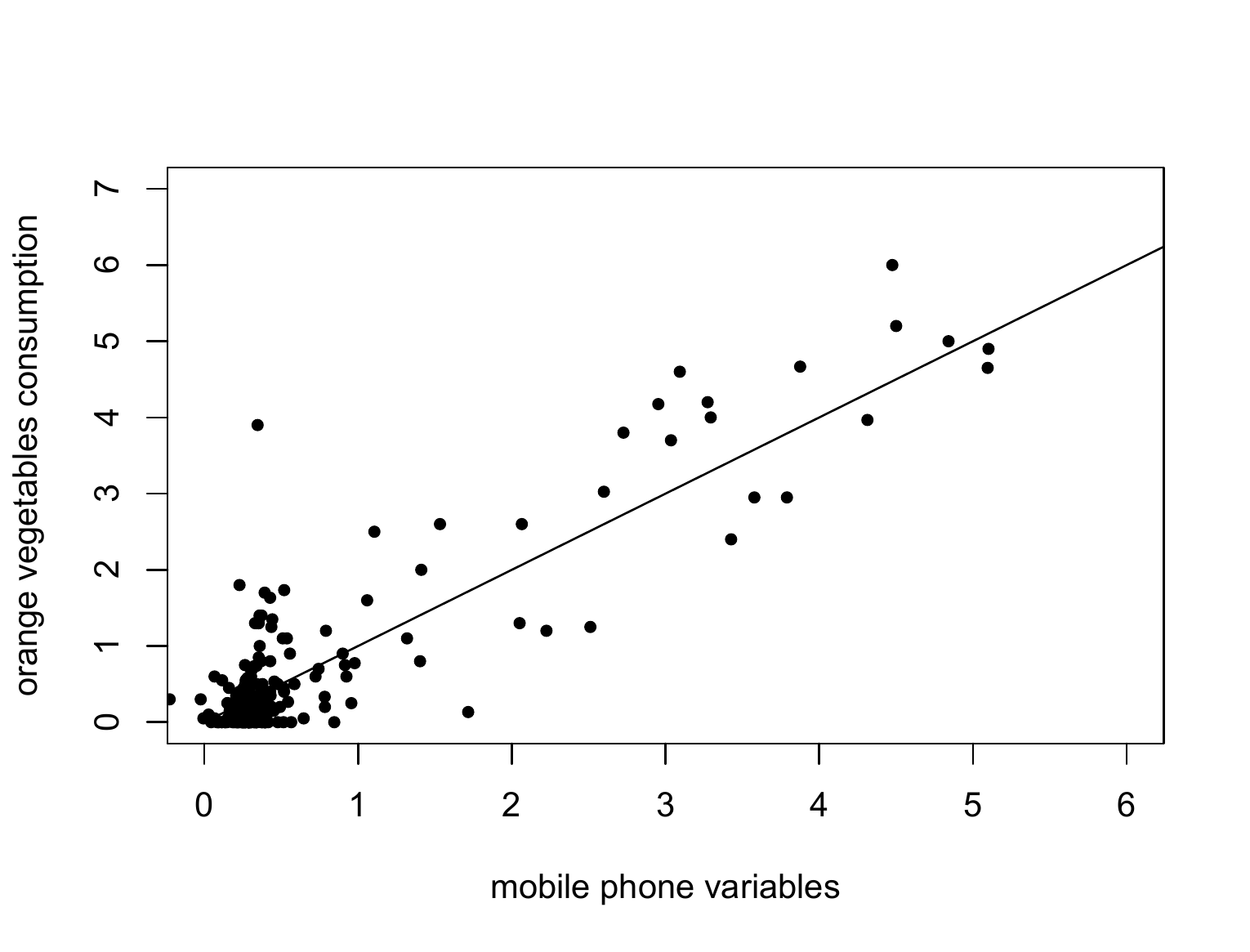}
\caption{{\bf Quadratic combination of CDR variables against carrots and orange vegetables consumption.} Correlation coefficient: 0.89.}
\label{combi_carrot}
\end{figure}

\begin{figure}[H]
\includegraphics[scale=0.6]{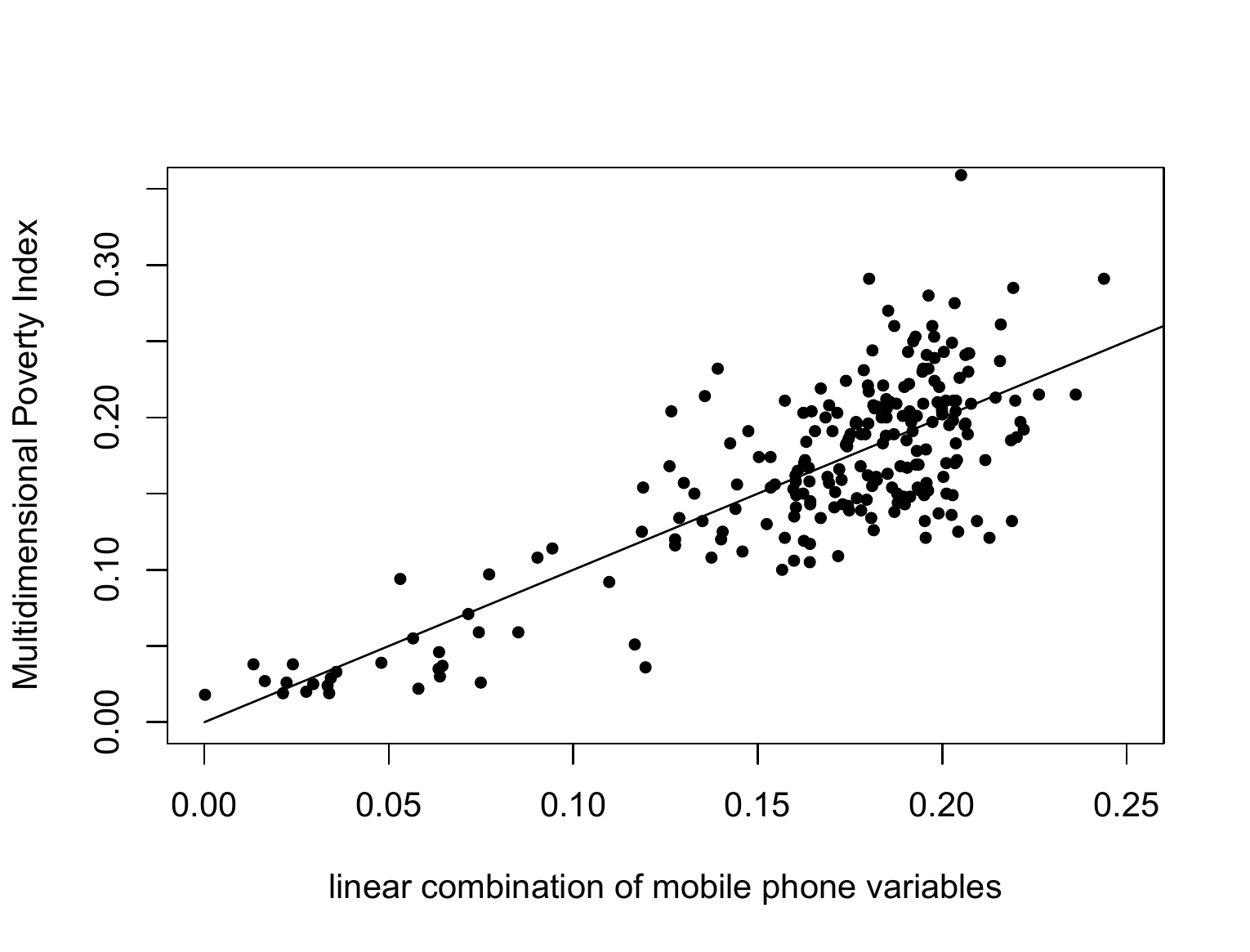}
\caption{{\bf Linear combination of CDR variables against MPI.} Correlation coefficient: 0.8.}
\label{combili_MPI}
\end{figure}

\section*{Tables}
% 
% See introductory notes if you wish to include sideways tables.
%
% NOTE: Please look over our table guidelines at http://www.plosone.org/static/figureGuidelines#tables to make sure that your tables meet our requirements. Certain types of spacing, cell merging, and other formatting tricks may have unintended results and will be returned for revision.
%
%\begin{table}[!ht]
%\caption{
%\bf{Table title}}
%\begin{tabular}{|c|c|c|}
%table information
%\end{tabular}
%\begin{flushleft}Table caption
%\end{flushleft}
%\label{tab:label}
% \end{table}
\begin{table}[H]
\caption{Correlation between the consumption of different types of food and the sum of airtime expenses.}
\begin{tabular}{| p{2cm} | p{9cm} | p{2cm} |}
\hline
\textbf{Correlation group } & \textbf{Name of variable (question: "how many times have you eaten [item] in the last 7 days?"  (answers between 0 and 7) )} & \textbf{Correlation with sum} \\
\hline
\hline
High & carrot, orange sweet potato (vitamin rich orange vegetables) & 0.82 \\
High & rice, wheat and other cereals &	0.76 \\
High & mandazi/chapatti/bread	& 0.76 \\
High & sugar and sweets &	0.70 \\
High & flesh meat &	0.69 \\
\hline 
\hline
Middle & eggs	& 0.59 \\
Middle & orange coloured fruits	& 0.57 \\
Middle & oil, fat, butter, ghee (including palm oil) &	0.54 \\
Middle & milk and milk products	& 0.50 \\
Middle & organ meat	& 0.48 \\
\hline 
\hline
Low & sorghum	& 0.43 \\
Low & ground nuts and seeds	& 0.37 \\
Low & other vegetables	& 0.35 \\
Low & fish	& 0.30 \\
Low & other fresh fruits	& 0.28 \\
Low & cooking banana	& 0.21 \\
Low & dark green leafy vegetables	& 0.18 \\
Low & beans, peas and other pulses	& 0.09 \\
Low & condiments	& 0.07 \\
Low & maize/ maize meal	& 0.04 \\
Low & other white roots and tubers	& 0.02 \\
Low & pumpkin, squash and other orange vegetables & 	0.01 \\
Low & cassava	  & -0.04 \\
\hline 
\hline
Negative & white sweet potato	& -0.41 \\
\hline
\end{tabular}
\begin{flushleft} The types of food are separated into four groups: those that correlate high, middle, and low with the sum of mobile phone expenses, and the white sweet potato forms the last group that correlates significantly negatively with the sum of expenses on mobile phones. 
\end{flushleft}
\label{table_food}
\end{table}

\begin{table}[H]
\caption{Selected variables from household surveys with high correlation with the sum of expenses in mobile consumption. }
\begin{tabular}{|l|l|l|}
\hline
\textbf{survey variable} & \textbf{cor. sum} & \textbf{$95\%$ conf.  int.}   \\
\hline
Share food from production & -0.64  &  $[-0.71, -0.55]$ \\
\hline
Share food from purchases & 0.66  & $[0.58, 0.73]$ \\
\hline
Food consumption score & 0.62  & $[0.54, 0.7]$ \\
\hline
Monthly food expenditure & 0.82   &   $[0.77, 0.86]$\\  
\hline
Education expenditure & 0.72   &   $ [0.65 , 0.78]$\\  
\hline
Communication expenditure & 0.69   &   $[0.62 , 0.75]$\\  
\hline
Total monthly expenditure & 0.6  & $[0.51, 0.68]$ \\ 
\hline
Per capita income & 0.63  & $[0.54, 0.7]$  \\ 
\hline
Multidimensional poverty index & -0.75   & $[-0.8, -0.69]$ \\ 
\hline
\end{tabular}
\begin{flushleft} 
The correlation shown here is the correlation between each indicator, and the sum of expenses on mobile phone airtime credit, along with their $95\%$ confidence intervals. All p-values are below $10^{-15}$. 
 \end{flushleft}
\label{table_cor}
\end{table}

%\section*{Supporting Information Legends}
%
% Please enter your Supporting Information captions below in the following format:
%\item{\bf Figure SX. Enter mandatory title here.} Enter optional descriptive information here.
% 
%\begin{description}
%\item {\bf}
%\item {\bf}
%\end{description}

\end{document}